\documentclass[prl,twocolumn,longbibliography,superscriptaddress, nobibnotes]{revtex4-2}

\usepackage[normalem]{ulem} 
\usepackage{graphicx} 
\usepackage[tbtags]{amsmath}
\usepackage{dsfont}
\usepackage{amsfonts}
\usepackage{hyperref}
\usepackage{siunitx}
\hypersetup{colorlinks=true, linkcolor=black, citecolor=black, urlcolor=blue}
\usepackage{orcidlink}
\usepackage[caption=false]{subfig} 
\usepackage{bm}

\begin{document}

\title{Tunable-Size Unruh--DeWitt Detector in a Multimode Cavity}

\author{Simon Brunner\,\orcidlink{0009-0007-6529-6331}}
\email{simon.brunner@uibk.ac.at}
\affiliation{Institut f\"ur Theoretische Physik, Universit{\"a}t Innsbruck, A-6020~Innsbruck, Austria}

\author{Farokh Mivehvar\,\orcidlink{0000-0003-4776-1352}}
\affiliation{Institut f\"ur Theoretische Physik, Universit{\"a}t Innsbruck, A-6020~Innsbruck, Austria}
\affiliation{Institute for Quantum Optics and Quantum Information, Austrian Academy of Sciences, A-6020 Innsbruck, Austria}
\affiliation{Many-Body Open Quantum Systems Unit, Okinawa Institute of Science and Technology Graduate University, Onna, Okinawa 904-0495, Japan}

\author{Helmut Ritsch\,\orcidlink{0000-0001-7013-5208}}
\affiliation{Institut f\"ur Theoretische Physik, Universit{\"a}t Innsbruck, A-6020~Innsbruck, Austria}

\author{Arkadiusz Kosior\,\orcidlink{0000-0002-5039-1789}}
\affiliation{Institut f\"ur Theoretische Physik, Universit{\"a}t Innsbruck, A-6020~Innsbruck, Austria}
\affiliation{Institute of Physics, Maria Curie-Skłodowska University, 20-031 Lublin, Poland}

\begin{abstract}
Quantum simulators based on Bose--Einstein condensates (BECs) provide a powerful platform to study relativistic quantum-field phenomena in the lab and in particular emulate relativistic analog particle detectors. However, such detectors typically couple to excitations beyond the acoustic phonon regime, so that the nonlinear Bogoliubov dispersion introduces corrections that cause the effective relativistic description to break down. We propose to overcome this limitation using an intrinsically momentum-selective Unruh--DeWitt detector model in which a tweezer-trapped atom couples to the condensate through a multimode optical cavity. Although the atom is effectively pointlike, the cavity-mediated interaction kernel causes it to sample density fluctuations over a controllable finite region of the BEC equivalent to a finite-size detector. The effective detector size is tunable via the cavity mode structure determining a controllable momentum cutoff that filters out nonphononic excitations. We analytically calculate the response of such a detector for uniform linear acceleration and uniform circular motion and show that this momentum selectivity can enhance the signal at experimentally accessible accelerations. Compatible with existing cavity-QED platforms our scheme thus provides a versatile setting for analog quantum-field measurements or relativistic quantum-information protocols.
\end{abstract}

\maketitle

\textit{Introduction.---}Local measurements in relativistic quantum field theory (QFT) are conceptually subtle because the theory admits no Lorentz-covariant particle-position operator \cite{hegerfeldtLocal}. Such measurements are therefore commonly described using localized quantum systems and, in particular, Unruh--DeWitt (UdW) particle detectors \cite{unruhNotesBlackholeEvaporation1976,takagiVacuumNoiseStress1986,2010-dewitt}. These are typically modeled as point-like two-level systems whose excitations probe field fluctuations \cite{crispinoUnruhEffectIts2008,takagiVacuumNoiseStress1986}. Detector models are a central concept used in relativistic quantum-information protocols, including entanglement harvesting, \cite{valentiniNonlocalCorrelationsQuantum1991,percheFullyRelativisticEntanglement2024,maeso-garciaEntanglementHarvestingState2022,pozas-kerstjensEntanglementHarvestingElectromagnetic2016,reznikEntanglementVacuum2003,reznikViolatingBellsInequalities2005,saltonAccelerationassistedEntanglementHarvesting2015,pozas-kerstjensHarvestingCorrelationsQuantum2015,saltonAccelerationassisted2015,clicheVacuumEntanglementEnhancement2011}, 
quantum communication, teleportation, and relativistic quantum computation \cite{alsingTeleportationUniformlyAccelerated2003,hottaQuantumMeasurementInformation2008, clicheRelativisticQuantumChannel2010,jonssonInformationTransmissionEnergy2015, funaiEngineeringNegativeStressenergy2017,tjoaQuantumTeleportationRelativistic2022,kasprzakTransmissionQuantumInformation2025,lemaitreUniversalQuantumComputer2025,asplingUniversalQuantumComputing2024,asplingDesignConstraintsUnruhDeWitt2024a}. Historically, the canonical application is the Unruh effect, in which a uniformly accelerated observer perceives the inertial vacuum as a thermal bath whose temperature is proportional to its acceleration  \cite{unruhNotesBlackholeEvaporation1976,takagiVacuumNoiseStress1986}.

Since the accelerations required for direct experimental tests are prohibitively large, in recent years numerous analog-gravity simulators have been proposed \cite{unruhExperimentalBlackHoleEvaporation1981,barceloAnalogueGravityBoseEinstein2001,heliumUniverse,braunsteinAnalogueSimulationsQuantum2023, schutzholdUltracoldAtomsQuantum2025a,barceloAnalogueGravity2026}. These include optical platforms \cite{adjeiQuantumSimulationUnruhDeWitt2020,yoonQuantumOpticalSimulator2026} and fermionic lattice systems \cite{rodriguez-lagunaSyntheticUnruhEffect2017,kosiorUnruhEffectInteracting2018,loukoThermalityRindlerQuench2018,kosiorNonlinearEntanglement2020}, but many are based on Bose--Einstein condensates (BECs) \cite{huQuantumSimulationUnruh2019,shengQuantumSimulationUnruh2021,percheBosePolaronsRelativistic2025,tianHarvestingEntanglementLorentzviolating2026,marinoCasimirForcesQuantum2017a,goodingInterferometricUnruhDetectors2020,retzkerMethodsDetectingAcceleration2008,fedichevGibbonsHawkingEffectSonic2003a,fedichevObserverDependencePhonon2004,comer2005superfluidanalogdaviesunruheffect}, where the approximately linear low-energy phonon dispersion constitutes a natural analog to simulate relativistic fields, also in curved spacetimes such as cosmological \cite{barceloProbingSemiclassicalAnalog2003,fedichevCosmologicalQuasiparticleProduction2004a,fischerQuantumSimulationCosmic2004a,jainAnalogModelFriedmannRobertsonWalker2007a,hungCosmologyColdAtoms2013,eckelRapidlyExpandingBoseEinstein2018,sparnExperimentalParticleProduction2024,viermannQuantumFieldSimulator2022a,schmidtEmergentspin1,tolosa-simeonCurvedExpandingSpacetime2022a,banikAccurateDeterminationHubble2022} and black-hole backgrounds \cite{garaySonicAnalogGravitational2000a,garaySonicBlackHoles2001,lahavRealizationSonicBlack2010a,kolobovObservationStationarySpontaneous2021,munozdenovaObservationThermalHawking2019,steinhauerObservationQuantumHawking2016,tajikExperimentalObservationCurved2023}. Obviously, accurately simulating a detector is as important as simulating the relativistic field itself. Existing BEC-based proposals model analog particle detectors using atomic quantum dots \cite{comer2005superfluidanalogdaviesunruheffect,fedichevGibbonsHawkingEffectSonic2003a,fedichevObserverDependencePhonon2004,retzkerMethodsDetectingAcceleration2008}, localized laser probes \cite{goodingInterferometricUnruhDetectors2020} or trapped Bose polarons \cite{percheBosePolaronsRelativistic2025,tianHarvestingEntanglementLorentzviolating2026,marinoCasimirForcesQuantum2017a}. In particular the implementations using trapped atoms appear as effectively pointlike detectors as the spatial extent of their wavefunctions is typically negligible compared with the relevant BEC length scales. Although mathematically convenient, this point-like approximation can lead to ultraviolet divergences, whereas finite-size detector models would provide natural regularization~\cite{martin-martinezGeneralRelativisticQuantum2020}. A further challenge arises at high accelerations which enlarges the detector response but also shifts its sensitivity towards shorter wavelengths beyond the acoustic regime, where the nonlinear Bogoliubov dispersion introduces corrections \cite{liberatiAnalogueQuantumGravity2006,weinfurtnerCosmologicalParticleProduction2009,chaProbingScaleInvariance2017,chandranExpansioncontractionDualityBreaking2025,schmidtSuperluminalModesQuantum2026} that deviate from the relativistic QFT model prediction.

In this Letter we show how to overcome this obstacle via a tunable-size UdW detector. Our setup combines a uniform BEC trapped in the transverse plane of a multimode optical cavity with a two-level particle tweezer-trapped at a distant antinode to implement the detector. A family of many near-degenerate transverse cavity modes mediates a nonlocal coupling with a finite transverse range that allows the point-like atom to sample density fluctuations over an extended region of the BEC. From the condensate’s perspective, the probe therefore behaves as an effective finite-size detector whose size is tunable through the cavity mode structure \cite{vaidyaTunableRangePhotonMediatedAtomic2018b,Mivehvar2021Cavity}. This tunability provides a controllable momentum cutoff that filters out non-phononic excitations. Moreover, spatially separating the probe atom from the condensate avoids the classical disturbances caused by dragging a probe through the BEC. In the following, we calculate the response for the two cases of a detector undergoing uniform linear acceleration along a one-dimensional (1D) trajectory and uniform circular motion in two dimensions (2D). Remarkably, the finite detector size can enhance the response to linear acceleration at experimentally accessible values because the detector samples a range of local proper accelerations across its spatial support. This enhancement makes the platform promising for analog quantum-field simulation studies.

\begin{figure}[t]
    \centering
    \includegraphics[width=\linewidth]{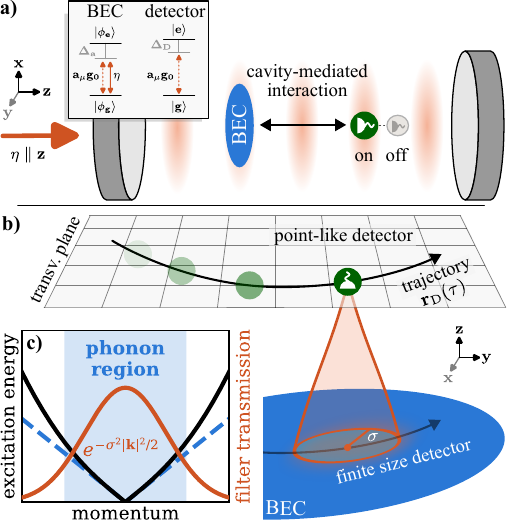}
    \caption{
    a)~Setup. A 1D/2D BEC confined in the transverse plane of a longitudinally pumped ($\eta$) near-planar optical cavity. A two-level atom (green circle), held by an optical tweezer outside the BEC, serves as an Unruh--DeWitt detector. Condensate and atom couple to many nearly degenerate cavity modes, which mediate an effective, finite-range interaction between them. b)~Effective model. The detector is steered along an accelerated trajectory in the transverse plane. Because the cavity-mediated coupling has a finite transverse range, the physically pointlike atom couples to an extended region of the condensate (orange cone), acquiring an effective detector size $\sigma$ (projected ellipse on the BEC). c)~Schematic momentum filtering. The black solid curve shows the Bogoliubov dispersion, which agrees with the relativistic linear dispersion (blue dashed line) only in the phononic regime $|\mathbf{k}|\xi\lesssim1$ indicated by the shaded region. The red curve shows the Gaussian transmission of the cavity-mediated detector profile for $\sigma \xi = 1.5$, which suppresses modes with momenta above the inverse detector size. Tuning the interaction range $\sigma$ controls the momentum cutoff and restricts the detector response to the acoustic, relativistic sector of the condensate fluctuations.}
    \label{fig:model}
\end{figure}

\textit{Setup and Effective Hamiltonian.}---We consider a longitudinally pumped, near-planar optical cavity containing a 1D (or 2D) BEC of alkali atoms and a spatially separated two-level probe atom as illustrated in Fig.~\ref{fig:model}(a).
The BEC is confined in the transverse plane at a common antinode of the longitudinal standing wave cavity modes, while the probe atom is held in optical tweezers at a position outside the BEC but within the cavity. Coupling both the BEC and the probe atom to a family of near-degenerate transverse cavity modes with single-photon Rabi frequency $g_0$ mediates an effective interaction between them via photon exchange.
The longitudinal standing-wave profile of the intracavity field creates uniform coupling between all antinodes within the cavity, whereas the transverse multi-mode structure confines interactions to a finite range perpendicular to the cavity axis due to interference of the individual mode contributions \cite{Salzburger2002Enhanced}. Increasing the number of participating modes narrows the radial coupling range. This cavity-mediated interaction allows a spatially separated probe to sample BEC density fluctuations over a controllable range and detect condensate phonons. 

The position of the probe atom plays two independent roles. First, moving the atom between a node and an antinode of the standing-wave envelope controls its coupling to the mode family and implements a switching function $\chi(t)$ as illustrated in Fig.~\ref{fig:model}(a). Second, moving the atom in the transverse $x$-$y$ plane couples it to different spatial regions of the BEC, thereby simulating a detector moving along an arbitrary trajectory, as depicted in Fig.~\ref{fig:model}(b).

In a frame rotating at the pump laser frequency, chosen to be in a far-detuned dispersive regime, we can adiabatically eliminate the internal excited states of the condensate atoms and the fast cavity modes. Assuming that contact interactions dominate over the cavity-induced nonlocal atom--atom interaction within the BEC we obtain the effective detector--condensate Hamiltonian
\begin{equation}
\begin{split}
    &\hat H = \int_{\mathbf{r}}  \hat\phi^\dagger\left[-\tfrac{\hbar^2}{2M}\nabla^2 + V_\mathrm{eff} +\tfrac{1}{2} U_0 \hat\phi^\dagger\hat\phi \right]\hat \phi - \frac{\hbar \Delta_\mathrm{D}}{2} \hat\sigma_z \\
    &+\hbar \lambda \chi(t)\left( \hat\sigma_+ + \hat \sigma_- \right)\int_{\mathbf{r}}   D(\mathbf{r}, \mathbf{r}_\mathrm{D}) \hat \phi^\dagger(\mathbf{r})\hat\phi(\mathbf{r}),
\end{split}
\label{eq:eff_hamiltonian}
\end{equation}
where $\int_\mathbf{r} = \int d^dr$ and $d$ is the effective dimension of the condensate. For a 2D BEC the integration is carried over the transverse plane, with $\mathbf r=(x,y)$, while for a 1D condensate  $\mathbf{r} = x$. Correspondingly, $\hat\phi(\mathbf r)$ denotes the condensate field operator. Here $M$ and $U_0$ denote the atomic mass and contact-interaction strength, while $\hat\sigma_z$ and $\hat\sigma_\pm$ are the detector's Pauli-$z$ and transition operators. The effective potential $V_\mathrm{eff}=V_\mathrm{ext}+\hbar\eta^2/\Delta_a$ combines the external trapping potential with the pump-induced light shift, where $\eta$ is the effective pump amplitude and $\Delta_a=\omega_p-\omega_a$ is the pump detuning from the condensate's electronic transition. Similarly, $\Delta_\mathrm{D}=\omega_p-\omega_\mathrm{D}$ is the pump detuning from the detector transition and sets the detector's energy gap. The effective detector--condensate coupling $\lambda=\eta g_0^2/(\Delta_a\Delta_{c,0})$ is controlled by the drive amplitude and the atomic and cavity detunings, where $\Delta_{c,0}=\omega_{c,0}-\omega_p$ denotes the detuning from the fundamental $\mathrm{TEM}_{00}$ cavity mode. A detailed derivation of Eq.~\eqref{eq:eff_hamiltonian} is given in the End Matter.

The second line of Eq.~\eqref{eq:eff_hamiltonian} describes the detector--condensate coupling. Owing to the finite transverse range of the cavity-mediated interaction, the detector samples the condensate density over a finite region $\sigma$ and couples to the smeared density $\hat n_\sigma(\mathbf r_\mathrm{D})\equiv\int_{\mathbf r}D(\mathbf r,\mathbf r_\mathrm{D})\hat\phi^\dagger(\mathbf r)\hat\phi(\mathbf r)$. This gives the pointlike detector an effective size set by the interaction range, see Fig.~\ref{fig:model}(b). The cavity mode structure determines the smearing kernel,
\begin{equation}\label{eq:D_kernel}
    D(\mathbf{r}, \mathbf{r}') = \frac{1}{\left( \sqrt{2\pi}\,\sigma \right)^d }
    e^{-|\mathbf{r} - \mathbf{r}'|^2/2\sigma^2},
\end{equation}
a normalized $d$-dimensional Gaussian of width $\sigma=\sqrt{w_0^2\epsilon/2}$, where $w_0$ is the waist of the fundamental mode. The width is governed by the dispersion of the transverse modes, which we assume to be exponential \cite{sm}. The small parameter $\epsilon=\delta L/\Delta_{c,0}$ quantifies the departure from ideal mode degeneracy, while $1/\epsilon$ estimates the number of transverse modes supported by the cavity. Tuning either the deviation $\delta L$ of the cavity length from the degenerate configuration or the cavity--pump detuning $\Delta_{c,0}$ therefore provides in situ control of the interaction range $\sigma$ \cite{vaidyaTunableRangePhotonMediatedAtomic2018b}.

We treat the detector--condensate coupling as a time-dependent perturbation. Before being switched on, the detector is in its ground state and sits on a node of the standing wave of the intracavity field while the atoms form a spatially homogeneous BEC with mean-field wavefunction $\phi_0=\sqrt{n_0}$, where $n_0$ is the atomic density.
Switching on the coupling, by moving the probe atom to an antinode, induces excitations that yield a nonzero detector response for accelerated detector motion. 

\textit{Elementary excitations of BEC and relativistic description.---}To describe the elementary excitations of the BEC, we expand the field operator around the homogeneous mean-field solution as $\hat{\phi} = \phi_0 + (\delta \hat{\phi}_\mathrm{re} + i\,\delta \hat{\phi}_\mathrm{im})/\sqrt{2}$. Keeping terms up to quadratic order in the fluctuations gives the standard Bogoliubov Hamiltonian. This Hamiltonian is diagonalized by a Bogoliubov transformation to quasiparticle modes $(\hat b_\mathbf{k},\hat b_\mathbf{k}^\dagger)$ \cite{sm}, yielding
\begin{equation}\label{disp_relation}
\hat H_\mathrm{at} = \sum_\mathbf{k\ne0} E_\mathbf{k} \hat b_\mathbf{k}^\dagger \hat b_\mathbf{k}, \quad 
E_\mathbf{k}=\hbar c |\mathbf{k}|\sqrt{1+\frac{|\mathbf{k}|^2\xi^2}{2}} .
\end{equation}
Here, $c=\sqrt{U_0n_0/M}$ is the speed of sound and $\xi=\hbar/\sqrt{2MU_0n_0}=\hbar/(\sqrt{2}Mc)$ is the healing length. The scale $\xi$ marks the crossover between collective phonons and free particle-like Bogoliubov excitations.

In order to simulate the analog of the relativistic Unruh effect, the detector should predominantly couple to modes for which the excitation spectrum is linear in $|\mathbf{k}|$. This is naturally satisfied in the phononic regime $|\mathbf{k}|\xi\ll1$, where $E_\mathbf{k}\simeq\hbar c|\mathbf{k}|$. In this limit, $\delta\hat \phi_\mathrm{im}$ admits an effective relativistic field-theory description, with the action
\begin{equation}
    S = \frac{\hbar^2}{4M}\int_{t, \mathbf{r}}\sqrt{g}\left[ g^{\mu\nu}  \partial_\mu (\delta\hat \phi_\mathrm{im})\partial_\nu (\delta\hat \phi_\mathrm{im}) \right],
\end{equation}
where the analog Lorentzian background geometry is given by the acoustic metric $g_{\mu\nu} = \mathrm{diag}(1, -1/c^{2}, -1/c^2)$ and the condensate's speed of sound $c^2(t, r) = U_0(t) n_0(r)/M$ \cite{tolosa-simeonCurvedExpandingSpacetime2022a}. In the above, the metric determinant $\sqrt{g}= |\det(g_{\mu\nu})|^{1/2} = 1/c^2$ guarantees a Lorentz-invariant integration measure.  

In our case, the speed of sound $c$ is constant, hence $g_{\mu \nu}$  can be brought to Minkowski form by the spatial rescaling $r\to cr$, i.e., $g_{\mu\nu}\to \mathrm{diag}(1,-1,-1)$. Moreover, in the phononic regime the relation
 \begin{equation}
   \delta\hat \phi_\mathrm{re}(t, \mathbf{r}) =-\frac{\hbar}{2M} \sqrt{g}~\partial_t (\delta\hat \phi_\mathrm{im})
\end{equation}
holds between the fluctuation fields. Thus $\delta\hat \phi_\mathrm{re}$ is proportional to the conjugate momentum of the emergent scalar field $\delta\hat \phi_\mathrm{im}$, with the canonical commutation relation $[\delta \hat\phi_\mathrm{re}(t, r'), \delta \hat\phi_\mathrm{im}(t, r)] = i \delta(r-r')$.

The cavity-mediated interaction couples the detector to the density of the BEC and therefore to its density fluctuations. Expanding $\hat\phi^\dagger\hat\phi$ to first order around the mean-field density gives a constant background term (which leads to a classical drive of the detector and can be neglected because it does not contribute to the excitation probability rate, as it produces only bounded Rabi oscillations and no long-time transition rate \cite{sm}), plus a fluctuation term proportional to $\delta\hat\phi_\mathrm{re}$. Hence, the corresponding interaction Hamiltonian in the interaction picture reads
 \begin{equation}
    H_I(t) = \hbar\tilde\lambda(t)\int_\mathbf{r} D(\mathbf{r}, \mathbf{r}_\mathrm{D})\delta\hat \phi_\mathrm{re}(t, \mathbf{r})  \left[ \hat\sigma_+(t)+ \mbox{H.c.}\right] ,
\end{equation}
where $\tilde \lambda(t) = \lambda\sqrt{2n_0} \chi(t)$ and $\hat\sigma_+(t) = \hat\sigma_+e^{-i\Delta_\mathrm{D}t}$. Note that in the limit of exact mode degeneracy $\sigma \rightarrow 0$, we obtain
\begin{equation}
    \lim_{\sigma \rightarrow 0} D(\mathbf{r}, \mathbf{r}_\mathrm{D})  =  \delta(\mathbf{r} - \mathbf{r}_\mathrm{D}), 
\end{equation}
and therefore our setup exactly realizes an idealized pointlike UdW detector that couples to the conjugate momentum of an emergent relativistic scalar field.

\textit{Tunable detector size.---}While in the pointlike limit we recover an ideal UdW detector at low energies, highly accelerated trajectories can also sample higher-momentum Bogoliubov modes where the dispersion is no longer relativistic, a well-known challenge in analog Unruh protocols. 
Crucially, our platform circumvents this limitation. The key advantage of our cavity-mediated implementation is that the interaction kernel $D(\mathbf r,\mathbf r_\mathrm{D})$ acts as an effective spatial smearing function of width $\sigma$, yielding an effective size for the point-like atomic detector. Equivalently, the detector couples to a Gaussian-smeared density fluctuation, whose Fourier transform multiplies each momentum component by
\begin{equation}
    \int D(\mathbf r,\mathbf r_\mathrm{D})e^{-i\mathbf k\cdot\mathbf r}\,d\mathbf r \propto e^{-\sigma^2|\mathbf k|^2/2}.
\end{equation}
Modes with $|\mathbf k|\gtrsim1/\sigma$ are therefore exponentially suppressed. Since the interaction range $\sigma=\sqrt{w_0^2\epsilon/2}$ is tunable through the cavity mode dispersion, our setup provides direct control over the detector's momentum cutoff. In particular, choosing an interaction range larger than the condensate healing length, $\sigma \gtrsim \xi$, isolates the acoustic sector by suppressing trans-Planckian modes for which the Bogoliubov spectrum deviates from a linear dispersion [see Fig.~\ref{fig:model}(c)]. 
As shown below, although this introduces finite frequency smearing into the response rate, the same filtering mechanism can be leveraged to substantially enhance the detectable analog signal.

\begin{figure*}[tbh]
    \centering
    \includegraphics[width=\linewidth]{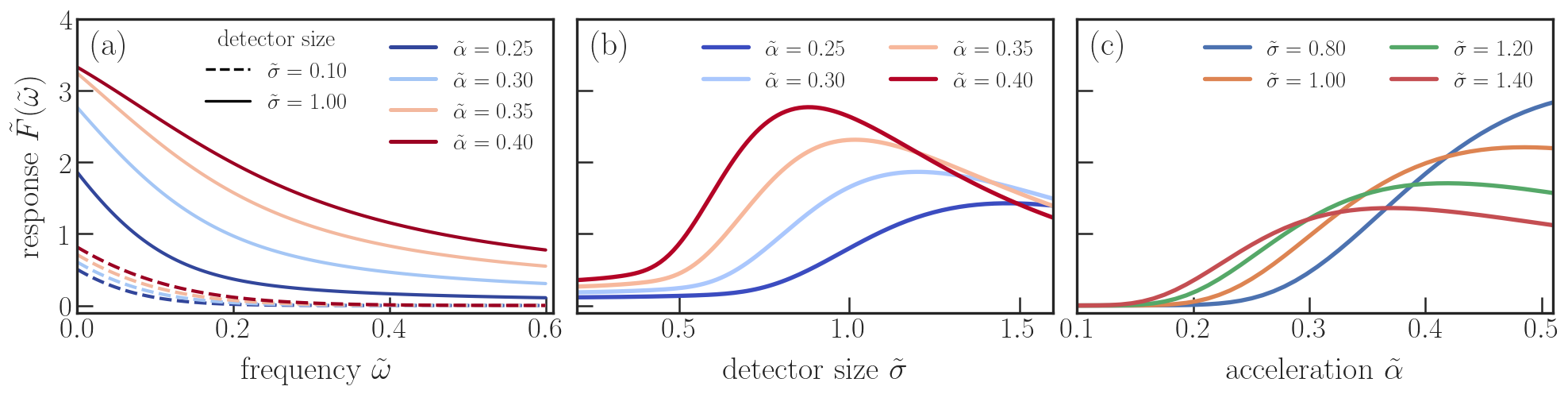}
    \caption{
    Detector response $\tilde{F}(\tilde{\omega})$ as a function of frequency, detector size, and acceleration. (a) Response versus dimensionless frequency $\tilde{\omega}$ for small ($\tilde{\sigma} = 0.10$, dashed lines) and large ($\tilde{\sigma} = 1.00$, solid lines) detector sizes across various accelerations $\tilde{\alpha}$. For $\tilde{\sigma} = 0.10$, the numerical results are indistinguishable from the analytical pointlike Unruh response formula. (b) Response as a function of detector size $\tilde{\sigma}$ for a fixed frequency $\tilde{\omega} = 0.1$, showcasing a distinct resonance-like amplification peak that shifts with acceleration. (c) Response versus acceleration $\tilde{\alpha}$ at a fixed frequency $\tilde{\omega} = 0.1$ for several large detector sizes, illustrating how finite-size effects optimize the signal at lower accelerations. } 
    \label{fig:numerical_response}
\end{figure*}

\textit{Response rate formula.}---In the Unruh effect, a detector responds only if it moves along a non-inertial worldline $\left( t(\tau), \mathbf{r}_\mathrm{D}(\tau) \right)$, which we  conventionally parameterize by its proper time $\tau$. Concretely, the response rate of the detector is defined as the excitation probability per interaction time, given by \cite{takagiVacuumNoiseStress1986,Louko2018,sm}:
\begin{equation}\label{rate}
 F(\omega
 ) = \lambda^2 \int_{-\infty}^{\infty} \mathrm{d}\tau \, e^{-i\omega\tau} 
 \chi(\tau) 
 \, g(\tau),
\end{equation}
which is the Fourier transform of the density-fluctuation correlator
$
g(\tau) = \left\langle \delta\hat{n}_{\sigma}(\tau) \, \delta\hat{n}_{\sigma}(0) \right\rangle,
$
with the smeared density fluctuation operator defined as
\begin{equation}
 \delta\hat{n}_{\sigma}(\tau) \equiv \int \mathrm{d}\mathbf{r} \, D\big(\mathbf{r}, \mathbf{r}_\mathrm{D}(\tau)\big) \, \delta\hat{n}\big(t(\tau), \mathbf{r}\big), 
\end{equation}
with $\delta \hat n=\sqrt{2n_0}\,\hat\varphi_\mathrm{re}$. 
Throughout, we consider only positive detector frequencies, $\omega=-\Delta_\mathrm{D}>0$, corresponding to a red-detuned detector transition.

In the phononic regime, where $E_\mathbf{k}\cong\hbar c|\mathbf{k}|$, the Gaussian detector profile gives 
\begin{align}
    g(\tau) = \frac{1}{V} \sum_\mathbf{k} e^{-\sigma^2 |\mathbf{k}|^2} \frac{E_\mathbf{k}}{2 U_\textrm{0}} e^{i  \left[ \mathbf{k} \cdot \mathbf{r}_\mathrm{D}(\tau) - E_\mathbf{k} t(\tau) / \hbar \right] }  
   \end{align}
with $\Delta\mathbf{r}_\mathrm{D}(\tau) =  \mathbf{r}_\mathrm{D}(\tau) -\mathbf{r}_\mathrm{D}(0)$. 
Furthermore, to simplify calculations we adopt dimensionless units based on the healing length $\xi$ and sound speed $c$, i.e.,  $\tilde \tau=\tau c/\xi$, $ \tilde{\mathbf{r}} = \mathbf{r}/\xi$, $ \tilde{\mathbf{k}}= \mathbf{k} \,\xi$, $\tilde\omega=\omega\xi/c$,  $\tilde\sigma=\sigma/\xi$, and define  the dimensionless rate function as 
\begin{align}
    \tilde F(\tilde \omega) = \frac{\sqrt{2}(2\pi)^d c\, \xi^{d-1} }{\lambda^2 n_0 }  
    F( \omega)= \int_{-\infty}^{\infty} d\tilde{\tau}  e^{-i \tilde{\omega} \tilde{\tau}}  \chi (\tilde \tau ) \tilde g(\tilde\tau),
\end{align}
with $ 
\tilde g(\tilde\tau)= \int_{-\infty}^{\infty} \mathrm{d} \tilde{\mathbf{k}} |\tilde{\mathbf{k}}|   e^{-\tilde{\sigma}^2 \tilde{\mathbf{k}}^2}  e^{i [ \tilde{\mathbf{k}}\cdot \Delta \tilde{\mathbf{r}}_\mathrm{D}(\tilde{\tau}) - |\tilde{\mathbf{k}}| \tilde{t}(\tilde\tau) ] }$ 
in thermodynamic limit.

\textit{Unruh effect in $(1+1)$ dimensions.}---To demonstrate the advantages of a setup with a tunable detector size, we begin by considering a detector uniformly accelerating along the $x$-direction in $1+1$ dimensions. Its worldline is parameterized by

\begin{equation}
\tilde t(\tilde\tau)
=
\frac{1}{\tilde\alpha}
\sinh\!\left(\tilde\alpha\,\tilde\tau\right),
\qquad
\tilde x(\tilde\tau)
=
\frac{1}{\tilde\alpha}
\cosh\!\left(\tilde\alpha\,\tilde\tau\right),
\end{equation}
with $\tilde\alpha=\alpha\xi/c^2$ being the dimensionless proper acceleration. A direct integration yields
\begin{equation}\label{g_tau}
\tilde g(\tilde \tau)  = 
\frac{1}{\tilde \sigma^2}
\left[
1-
\sum_{s=\pm}
a_s
\left(
D(a_s)
+\frac{i\sqrt{\pi}}{2}e^{-a_s^2}
\right)
\right]
\end{equation}
with $a_\pm = [\tilde t(\tilde  \tau)\pm\Delta \tilde x(\tilde \tau)]/2\tilde \sigma$, and $D(a)$
being the Dawson function.
In the pointlike limit $\tilde \sigma\to0$,  we successfully recover the standard analytical UdW response with conjugate momentum coupling 
\begin{equation}
\tilde{F}_{\mathrm{th}}(\tilde{\omega}) = 
\lim_{\tilde \sigma \rightarrow 0} \tilde F (\tilde \omega)   = 
4\pi \frac{\tilde{\omega}}{\exp\left(  \tilde{\omega} / \tilde T_U  \right)  -1 },
\end{equation}
where the dimensionless Unruh temperature 
is $\tilde T_U = \frac{\tilde \alpha}{2\pi}$.

For a non-zero smearing width ($\tilde{\sigma} > 0$), we evaluate the Fourier transform of $\tilde{g}(\tilde{\tau})$ numerically, with the corresponding behavior presented across the three panels of Fig.~\ref{fig:numerical_response}. We find that for a small smearing width ($\tilde{\sigma} \lesssim  0.10$), the numerical results are virtually indistinguishable from the analytical Bose-Einstein distribution obtained in the ideal pointlike limit. However, as the detector size increases, the response visibly deviates from this standard distribution. Remarkably, rather than suppressing the signal, a finite detector size enhances the response amplitude. To explore this enhancement further, panels (b) and (c) analyze the response at a fixed frequency 
$\tilde{\omega} =-\Delta_D \xi/c = 0.1$  . Specifically, as a function of detector size $\tilde{\sigma}$ [panel (b)] and acceleration $\tilde{\alpha}$ [panel (c)], we observe that the maximum signal amplification shifts systematically. Consequently, increasing the detector size to $\tilde{\sigma} \sim 1$ allows us to strongly amplify the response in regimes of low acceleration, shifting the peak sensitivity toward smaller $\tilde{\alpha}$ values where the standard, pointlike Unruh effect would otherwise be deeply suppressed and challenging to detect.

The observed enhancement of the finite-size UdW detector response at low accelerations suggests a geometric interpretation based on Born rigidity under relativistic acceleration. A spatially extended detector can maintain its volume only if one of its edges accelerates faster than the other. Consequently, the enhancement could be understood as a contribution from the detector's finite spatial support extending into regions of higher proper acceleration, and hence, higher local Unruh temperature.

\textit{Circular motion Unruh effect in $(2+1)$ dimensions.}---In this section we show that our framework can be easily extended to more complex trajectories, such as a detector undergoing uniform circular motion in $(2+1)$ dimensions \cite{Biermann2020,Louko2018,DBunney2023,Parry2025,parryWaitingUnruh2026}, whose bounded, periodic trajectory is particularly suitable for analog implementations because it enables long interaction times and, more importantly, simplifies the implementation of analog time dilation through its constant Lorentz factor (see the End Matter discussion).

For this setup, the detector's worldline, parameterized by the proper time $\tilde{\tau}$, is given by
\begin{equation}
\left( \tilde{t}(\tilde{\tau}), \tilde{\mathbf{r}}(\tilde{\tau})\right) = \left(\gamma \tilde{\tau}, \tilde{R} \cos (\gamma \tilde{\Omega} \tilde{\tau}), \tilde{R} \sin (\gamma \tilde{\Omega} \tilde{\tau})\right),
\end{equation}
where $\tilde{R} = R / \xi$ is the dimensionless orbital radius, $\tilde{\Omega} = \Omega \xi/c$ is the dimensionless angular velocity, and $\gamma = 1/({1 - (\tilde{R} \tilde{\Omega})^2})^{1/2}$ is the Lorentz factor connecting coordinate time to proper time.

Unlike uniform linear acceleration, circular motion does not produce an exactly thermal Unruh spectrum. The detector response is stationary and nonzero, but it is nonthermal: the periodic trajectory resolves the field into discrete angular harmonics rather than a Planck distribution. In this case, the response can be calculated analytically even for a finite detector size $\tilde{\sigma}$, following standard treatments of detector response in circular motion \cite{Louko2018,Biermann2020,DBunney2023,Parry2025}. The result is a discrete sum over non-negative resonant momentum modes,
\begin{equation}\label{eq:circular_response}
\tilde{F}(\tilde{\omega}) = \frac{4\pi^2}{\gamma} \sum_{n: \, \tilde{k}_n \geq 0} \tilde{k}_n^2 e^{-\tilde{\sigma}^2 \tilde{k}_n^2} J_n^2\left(\tilde{R} \tilde{k}_n\right),
\end{equation}
where $J_n$ denotes the $n$-th order Bessel function of the first kind and $\tilde{k}_n = {(n\gamma \tilde{\Omega}- \tilde{\omega})}/{ \gamma}$.

Eq.~\eqref{eq:circular_response} also clarifies why the finite-size enhancement found for linear acceleration does not directly carry over to circular motion. For linear acceleration, the Born-rigidity intuition suggests that an extended detector samples different proper accelerations along the acceleration direction. For circular motion at fixed radius, by contrast, the acceleration is radial and there is no analogous extension toward a Rindler horizon. The detector size appears only through the damping factor $e^{-\tilde{\sigma}^2\tilde{k}_n^2}$, so increasing $\tilde{\sigma}$ simply filters out high-momentum harmonics rather than amplifying the response.

\begin{figure}[t]
    \centering
    \includegraphics[width=0.99\linewidth]{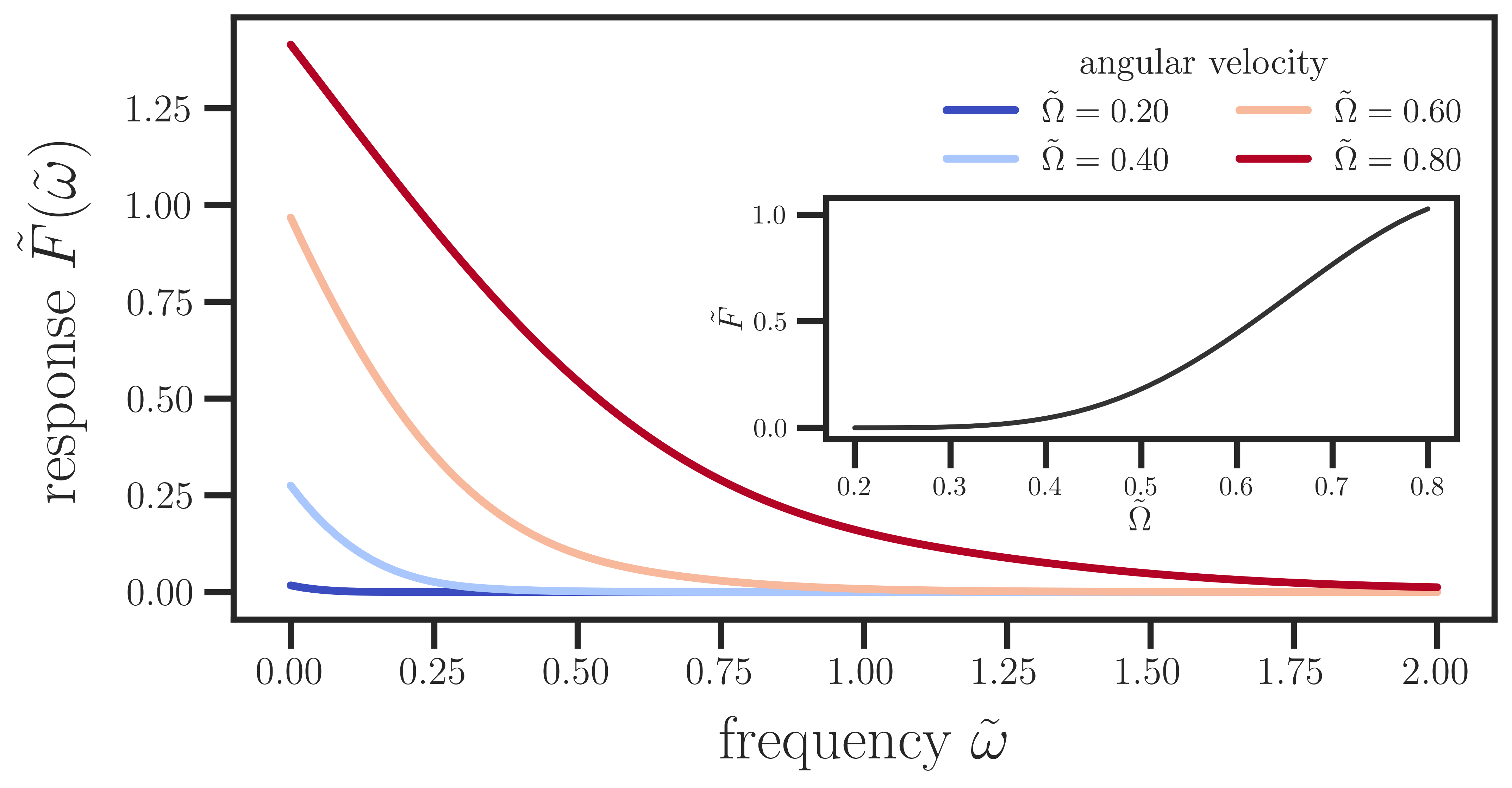}
\caption{Nonthermal detector response $\tilde{F}(\tilde{\omega})$ for circular motion in $(2+1)$ dimensions. The main panel shows the dimensionless response as a function of detector frequency for fixed detector size $\tilde{\sigma}=1$ and orbital radius $\tilde{R}=1$, with the four curves corresponding to $\tilde{\Omega}=0.20$, $0.40$, $0.60$, and $0.80$ (from blue to red). The response increases with angular velocity and decreases with detector frequency. Unlike uniform linear acceleration, circular motion does not yield an exactly thermal spectrum. \textit{Inset:} Response at fixed frequency $\tilde{\omega}=0.2$ as a function of angular velocity, showing monotonic growth toward the relativistic regime.}
    \label{fig:circular_response}
\end{figure}

We plot this nonthermal analytical response in Fig.~\ref{fig:circular_response} for $\tilde{\sigma}=\tilde{R}=1$. As illustrated in the main panel, the response $\tilde{F}(\tilde{\omega})$ decreases rapidly with increasing detector frequency $\tilde{\omega}$. Increasing the angular velocity $\tilde{\Omega}$ enhances the response across the plotted frequency range because more resonant harmonics satisfy $\tilde{k}_n>0$ and the circulating detector samples stronger vacuum fluctuations. The inset shows the same trend at fixed frequency $\tilde{\omega}=0.2$, where the response grows nonlinearly as the motion approaches the relativistic regime.

\textit{Conclusion.---}We introduced a novel cavity mediated non-contact detector for the density fluctuations of a trapped quantum gas. As genuine application we study its interpretation as 
an Unruh--DeWitt detector for a BEC in a multimode cavity. By choice of an appropriate mode structure we can transform a pointlike trapped detector atom to an effective, tunable size sensor with an inherent momentum filter. For the case of analog gravity simulations this restricts its response to the phononic (i.e. relativistic) sector of the BEC fluctuations and enhances the response signal at experimentally accessible accelerations. A further advantage of using a multi-mode cavity setup is that the detector state can be reconstructed non-destructively from the cavity output field, while the detector gap remains tunable via the pump-detector detuning. Beyond our chosen example the presented platform can be generalized to several detector atoms and presents itself as a versatile testbed for other protocols such as entanglement harvesting and decoherence measurements \cite{gundhiMeasuringDecoherenceDue2025}.


\begin{acknowledgments}
The authors thank Clemens Jakubec and Yaser Tavakoli for fruitful discussions. This research was funded in whole or in part by the Austrian Science Fund (FWF) [Grant DOIs: 10.55776/P35891, 10.55776/PAT2450125, and 10.55776/I6673]. For open access purposes, the author has applied a CC BY public copyright license to any author accepted manuscript version arising from this submission. AK acknowledges the support of the Polish National Agency for Academic Exchange (NAWA) under the ``Polish Returns'' programme, grant no. BPN/PPO/2024/1/00010/U/00001. 
\end{acknowledgments}

\textit{Data availability statement.}---The findings of this study are analytical and can be completely reconstructed using the formulas and derivations provided within the manuscript.

\bibliography{udw_cavity}

\onecolumngrid
\begin{center}
\vspace{0.8em}
{\large\bfseries END MATTER}
\vspace{0.5em}
\end{center}
\twocolumngrid

\textit{Experimental feasibility.}---All aspects of our setup are within reach of current experimental capabilites. A $^{87}$Rb condensate coupled to a multimode cavity has already been realized~\cite{vaidyaTunableRangePhotonMediatedAtomic2018b,guoOpticalLatticeSound2021}, with a minimum cavity-mediated interaction range of $\sigma\approx\SI{2}{\micro m}$, limited by mirror imperfections. This range, which is independent of the atomic species, sets the effective detector size and can be tuned in situ without disturbing the condensate. Compared against the typical healing length of a 2D $^{87}$Rb condensate, $\xi\approx \SI{0.3}{\micro m}-\SI{0.6}{\micro m}$ \cite{villeSoundPropagationUniform2018,sunamiObservationBerezinskiiKosterlitzThoulessTransition2022}, this places the dimensionless detector size $\tilde\sigma=\sigma/\xi$ in the finite-size regime ($\tilde\sigma>1$), where the response is enhanced (Fig.~\ref{fig:numerical_response}). 

The complementary pointlike regime ($\tilde\sigma<1$) can be reached by employing a $^{39}$K condensate where broad Feshbach resonances allow the tuning of the healing length over a large range, so that at fixed $\sigma$ both $\tilde\sigma>1$ and $\tilde\sigma<1$ become accessible within a single experimental platform \cite{fattoriAtomInterferometryWeakly2008}. The same interaction tunability already underpins existing analogue-cosmology experiments ~\cite{viermannQuantumFieldSimulator2022a}.

The detector is a single atom held in an optical tweezer. Such tweezers position atoms inside a cavity with nanometer-scale precision~\cite{seubertTweezerSubwave} and steer them along programmable trajectories with accelerations of up to $\SI{e4}{m\,s^{-2}}$~\cite{Hwang:23,hwangFastReliableAtom2025,bluvsteinQuantumProcessorBased2022}. Since the speed of sound in a typical BEC is only $c \sim \SI{1}{mm\,s^{-1}}$, the regime of interest $\tilde\alpha = \alpha \xi/c^2 \sim 1$ is reached at a laboratory acceleration of just $\alpha\lesssim\SI{1}{m\,s^{-2}}$. This acceleration is four orders of magnitude below what is currently possible.

\emph{Analog time dilation.---}%
Operating in this relativistic regime requires care in relating laboratory quantities to those of the analog spacetime. The physical detector gap is set by the detuning $\Delta_\mathrm{D}$ in the laboratory frame, whereas the rate \eqref{rate} involves the gap in the detector's rest frame \cite{sm}, $\omega = -\gamma(\tau)\,\Delta_\mathrm{D}$, with $\gamma = \mathrm{d}t/\mathrm{d}\tau$ the Lorentz  factor connecting lab time to analog proper time. A constant gap in \eqref{rate} therefore requires simulating analog time dilation, i.e.\ changing the detuning as $\Delta_\mathrm{D}(\tau) = \Delta_\mathrm{D}/\gamma(\tau)$ along the trajectory to have a one-to-one correspondence with the relativistic Unruh effect. 

This is precisely what makes circular motion attractive for analog experiments. Its Lorentz factor is constant along the trajectory, so a static laboratory detuning $\Delta_\mathrm{D}$ already realizes a constant gap $\omega = -\gamma\Delta_\mathrm{D}$ in the detector's rest frame. 

Operating in a regime with large $\gamma$ offers another advantage, since \eqref{rate} is a rate per unit
analog proper time while an experiment counts excitations per unit laboratory time. The two are related by the gamma factor $F_\mathrm{lab} = \gamma F$, which boosts the laboratory-frame signal as the detector velocity approaches the speed of sound and $\gamma$ grows.

\subsection{Microscopic Description}

We consider $N$ bosonic two-level atoms of transition frequency $\omega_a$ that form a BEC inside a near-planar multimode optical cavity, together with a single tweezer-trapped two-level atom of transition frequency $\omega_\mathrm{D}$, which serves as the detector. Near-planar resonators support families of many nearly degenerate transverse modes. Within a single such family, the mode functions factorize into transverse Hermite--Gaussian profiles $u_\mu(\mathbf{r})=u_l(x)u_m(y)$, labeled by the multi-index $\mu=(l,m)$, and a common longitudinal standing-wave envelope, $\mathcal{U}_\mu(\mathbf{r}, z) = u_\mu(\mathbf{r})\cos(kz)$, where $\mathbf{r}=(x,y)$ and $z$ denote the transverse and longitudinal coordinates, respectively. We denote the photon annihilation operator of mode $\mu$ by $\hat a_\mu$ and its resonance frequency by $\omega_{c,\mu}$. The condensate is confined to the transverse plane, with its $z$-position fixed to an antinode of the standing wave such that $\mathcal{U}_\mu(\mathbf{r}, z_\mathrm{BEC}) = u_\mu(\mathbf{r})$.

The detector atom is held at position $(\mathbf{r}_\mathrm{D}, z_\mathrm{D})$ and its longitudinal position defines the switching function $\chi(t) = \cos(k z_\mathrm{D}(t))$. Thus, the local mode amplitude at the detector factorizes as $\mathcal{U}_\mu(\mathbf{r}_\mathrm{D}, z_\mathrm{D}) = u_\mu(\mathbf{r}_\mathrm{D})\, \chi(z_\mathrm{D}(t))$.

The system is driven by a laser of frequency $\omega_p$ that pumps the fundamental [$\mu=(0,0)$] cavity mode with amplitude $\eta_0$, while all other modes remain unpumped. The extent of the BEC is assumed to be much smaller than the mode waist $w_0$, so that the fundamental mode amplitude is approximately uniform across the condensate, $u_0(\mathbf{r})\approx u_0$.

\textit{Microscopic Hamiltonian.}---We describe the condensate atoms by the bosonic field operators $\hat\phi_g$ and $\hat\phi_e$ for their internal ground and excited state, respectively, while the Pauli operators $\hat\sigma_z$ and $\hat\sigma_\pm$ act on the internal levels of the detector. In the rotating frame of the pump, the microscopic Hamiltonian reads
\begin{equation}\label{eq:full_hamiltonian}
\begin{split}
    &\hat H = \int_{\mathbf{r}} \hat\phi_g^\dagger\left[-\frac{\hbar^2}{2M}\nabla^2 + V_\mathrm{ext} +\frac{1}{2} U_0 \hat\phi_g^\dagger\hat\phi_g \right]\hat\phi_g \\
    &- \hbar\Delta_a \int_{\mathbf{r}} \hat\phi_e^\dagger\hat\phi_e + \hbar g_0 \sum_\mu \int_{\mathbf{r}} \left[ u_\mu(\mathbf{r})\, \hat\phi_e^\dagger \hat a_\mu \hat\phi_g + \mathrm{H.c.}\right] \\
    &- \hbar\sum_\mu \left[\Delta_{c,\mu}\, \hat a_\mu^\dagger \hat a_\mu + i\eta_0 \left(\hat a_0 - \hat a_0^\dagger\right)\right] - \frac{\hbar\Delta_\mathrm{D}}{2}\hat\sigma_z \\
    &+ \hbar g_0 \chi(t) \sum_\mu \left[ u_\mu(\mathbf{r}_\mathrm{D})\, \hat\sigma_+ \hat a_\mu + \mathrm{H.c.}\right],
\end{split}
\end{equation}
where $\int_{\mathbf{r}}\equiv\int d^dr$, $M$ is the atomic mass, $V_\mathrm{ext}$ the external trapping potential, and $U_0$ the contact-interaction strength. All frequencies are measured relative to the pump. $\Delta_a = \omega_p-\omega_a$ is the atomic detuning, $\Delta_\mathrm{D} = \omega_p - \omega_\mathrm{D}$ the detector detuning, and $\Delta_{c,\mu} = \omega_{c,\mu} - \omega_p$ the detuning of cavity mode $\mu$. Anticipating the far-detuned regime in which $\Delta_a$ is the dominant frequency scale of the dynamics, we have dropped the kinetic, trapping, and collisional terms of the excited state. Cavity losses enter through the replacement $\Delta_{c,\mu} \to \tilde{\Delta}_{c,\mu} = \Delta_{c,\mu} - i\kappa$. Here, we work in the dispersive regime, where the cavity decay rate $\kappa$ is negligible compared with the mode detunings, $\Delta_{c,\mu} \gg \kappa$. 

\textit{Elimination of the excited state.}---In the far-detuned dispersive limit the excited state remains essentially unpopulated and follows the slower degrees of freedom adiabatically. Setting $i\hbar\,\partial_t\hat\phi_e \stackrel{!}{=} 0$ in its Heisenberg equation of motion yields $\hat \phi_e = \frac{g_0}{\Delta_a} \sum_\mu u_\mu(\mathbf{r})\, \hat a_\mu \hat \phi_g.$
Substituting this back into Eq.~\eqref{eq:full_hamiltonian} eliminates the excited state in favor of a cavity-induced potential for the ground-state field, for which we write $\hat\phi \equiv \hat\phi_g$ from now on. This yields the following Hamiltonian
\begin{equation}\label{eq:H_ground}
\begin{split}
    &\hat H = \int_{\mathbf{r}} \hat\phi^\dagger\bigg[-\frac{\hbar^2}{2M}\nabla^2 + V_\mathrm{ext} +\frac{1}{2} U_0 \hat\phi^\dagger\hat\phi \\
    &\qquad\qquad + \frac{\hbar g_0^2}{\Delta_a}\sum_{\mu,\nu} u_\mu(\mathbf{r})\, u_\nu(\mathbf{r})\, \hat a_\mu^\dagger \hat a_\nu \bigg]\hat\phi \\
    &- \hbar\sum_\mu \left[\Delta_{c,\mu}\, \hat a_\mu^\dagger \hat a_\mu + i\eta_0 \left(\hat a_0 - \hat a_0^\dagger\right)\right] - \frac{\hbar\Delta_\mathrm{D}}{2}\hat\sigma_z \\
    &+ \hbar g_0 \chi(t) \sum_\mu \left[ u_\mu(\mathbf{r}_\mathrm{D})\, \hat\sigma_+ \hat a_\mu + \mathrm{H.c.}\right].
\end{split}
\end{equation}

\textit{Displacement of the pumped mode - effective drive.}---The fundamental mode is driven strongly and we expand its operator around the coherent steady-state amplitude, $\hat a_0 \to \langle\hat a_0\rangle + \hat a_0$, where $\hat a_0$ henceforth denotes the fluctuation. The Heisenberg equations of motion for the cavity modes read
\begin{equation}\label{eq:heisenberg_cavity}
\begin{split}
    i \partial_t \hat a_\mu ={}& -\Delta_{c, \mu}\, \hat a_\mu + i\eta_0 \delta_{\mu 0} + g_0\, \chi(t)\, u_\mu^*(\mathbf{r}_\mathrm{D})\,\hat{\sigma}_- \\
    &+ \frac{g_0^2}{\Delta_a}\sum_{\nu}\left( \int_\mathbf{r} u_\mu u_\nu\, \hat \phi^\dagger \hat \phi \right) \hat{a}_\nu ,
\end{split}
\end{equation}
which for the strongly and time-independently pumped fundamental mode give, to leading order, $\langle\hat a_0\rangle = i \eta_0/\Delta_{c,0}$. Inserting the displacement into Eq.~\eqref{eq:H_ground} and defining the effective condensate drive $\eta = \eta_0\, g_0\, u_0/\Delta_{c,0}$ brings the cavity-induced potential to the form $\hbar\,|i\eta + g_0\sum_\mu \hat a_\mu u_\mu(\mathbf{r})|^2/\Delta_a$. The displacement additionally generates the term $\hbar g_0 (\eta_0/\Delta_{c,0})\, u_0(\mathbf{r}_\mathrm{D})\,\chi(t) \left( i \hat{\sigma}_+ - i \hat{\sigma}_-\right)$, a classical drive of the detector, which does not contribute to the transition rate as it produces only bounded Rabi oscillations and no long-time transition rate \cite{sm}. We thus arrive at the Hamiltonian
\begin{equation}\label{eq:eff_pump_hamiltonian}
\begin{split}
    \hat H ={}&  \int_{\mathbf{r}} \hat\phi^\dagger\left[-\frac{\hbar^2}{2M}\nabla^2 + V_\mathrm{ext} +\frac{1}{2} U_0 \hat\phi^\dagger\hat\phi \right]\hat\phi \\
    &+ \frac{\hbar}{\Delta_a}\int_{\mathbf{r}}  \hat \phi^\dagger\, \Big|i\eta +g_0 \sum\nolimits_\mu\hat a_\mu u_\mu(\mathbf{r})\Big|^2\, \hat \phi \\
     &-\hbar \sum_\mu \Delta_{c, \mu} \hat a^\dagger_\mu \hat a_\mu - \frac{\hbar \Delta_\mathrm{D}}{2} \hat\sigma_z\\
& + \hbar g_{0}\chi(t)\sum_\mu \left[u_\mu(\mathbf{r}_\mathrm{D})\,\hat \sigma_+ \hat a_\mu + \mathrm{H.c.}\right].
\end{split}
\end{equation}

\textit{Elimination of the cavity modes.}---The cavity modes evolve on the fastest timescale, set by the detunings $\Delta_{c,\mu}$, and can therefore be adiabatically eliminated as well. From Eq.~\eqref{eq:heisenberg_cavity} their steady state is
\begin{equation} \label{eq:cavity_modes}
    \hat a_\mu = \frac{g_{0}}{\Delta_{c,\mu}}\left[\frac{i\eta}{\Delta_a}\int_{\mathbf{r}}u_\mu^*(\mathbf{r})\, \hat{\phi}^\dagger \hat\phi + \chi(t)\, u^*_\mu(\mathbf{r}_\mathrm{D})\, \hat\sigma_- \right].
\end{equation}
Substituting this expression back into Eq.~\eqref{eq:eff_pump_hamiltonian} yields
\begin{equation}\label{eq:H_nonlocal}
\begin{split}
    &\hat H = \int_{\mathbf{r}}  \hat\phi^\dagger\left[-\frac{\hbar^2}{2M}\nabla^2 + V_\mathrm{eff} +\frac{1}{2} U_0 \hat\phi^\dagger\hat\phi \right]\hat \phi - \frac{\hbar \Delta_\mathrm{D}}{2} \hat\sigma_z \\
    &+\hbar U_\mathrm{nl}\int_{\mathbf{r}}\int_{\mathbf{r}'} D(\mathbf{r}, \mathbf{r}')\,\hat \phi^\dagger(\mathbf{r}) \hat \phi(\mathbf{r})\, \hat \phi^\dagger (\mathbf{r}')\hat \phi(\mathbf{r}') \\
    &+\hbar \lambda \chi(t) \left( i\hat\sigma_+  + \mathrm{H.c.} \right)\int_\mathbf{r}    D(\mathbf{r}, \mathbf{r}_\mathrm{D})\,\hat \phi^\dagger(\mathbf{r})\hat\phi(\mathbf{r}) .
\end{split}
\end{equation}
Here $V_\mathrm{eff}=V_\mathrm{ext}+\hbar\eta^2/\Delta_a$ combines the external trapping potential with the pump-induced light shift, and the interaction kernel
\begin{equation}\label{eq:D_modesum}
    D(\mathbf{r}, \mathbf{r}') = \sum_\mu \frac{u_\mu(\mathbf{r})\, u_\mu(\mathbf{r}')}{\Delta_{c,\mu}/\Delta_{c,0}}
\end{equation}
is fixed by the transverse mode structure of the cavity. For a near-degenerate cavity, with an exponential mode dispersion $\Delta_{c,\mu} = \Delta_c\, e^{n_\mu\epsilon}$, where $n_\mu = l + m$, it reduces to the normalized Gaussian \eqref{eq:D_kernel} of the main text with width $\sigma=\sqrt{w_0^2\epsilon/2}$ \cite{sm}. The effective couplings are the detector--condensate interaction $\lambda = \eta g_0^2/\Delta_a\Delta_{c,0}$ and the cavity-induced nonlocal atom--atom interaction $U_\mathrm{nl} = (\eta g_0)^2/\Delta_a^2\Delta_{c,0}$.

In performing the substitution we have dropped the potential $\propto \sum_{\mu,\nu}\hat a_\mu^\dagger \hat a_\nu$ of Eq.~\eqref{eq:H_ground}, which only generates interaction terms of order $\Delta_a^{-3}$. Moreover, the second term of Eq.~\eqref{eq:cavity_modes} produces a detector self-interaction $\propto \hat\sigma_+\hat\sigma_-$ which amounts to a deterministic Stark shift of the detector gap. For a switching function that is constant during the interaction window this shift is constant, and we absorb it into the definition of $\Delta_\mathrm{D}$.

\textit{Effective model.}---We focus on the regime where the contact interaction dominates the cavity-induced nonlocal atom--atom interaction, $U_0 \gg U_\mathrm{nl}$, so that the second line of Eq.~\eqref{eq:H_nonlocal} can be neglected. Importantly, in this regime the detector coupling is still relevant. The ratio $\lambda/U_\mathrm{nl} = \Delta_a/\eta \gg 1$ is large because the atomic detuning is the dominant frequency scale in the far-detuned dispersive regime, so the detector--condensate interaction remains relevant even where the nonlocal condensate interaction is negligible. Dropping the nonlocal interaction and absorbing the residual phase into the detector operators, $\hat\sigma_\pm \to \mp i\,\hat\sigma_\pm$, yields the effective Hamiltonian of the main text.

\newpage
\clearpage
\clearpage
\onecolumngrid

\begin{center}
  {\large\textbf{Supplemental Material}\\[0.4cm]}
\end{center}

\setcounter{section}{0}
\setcounter{equation}{0}
\setcounter{figure}{0}
\setcounter{table}{0}

\section{Cavity-Mediated Interaction} \label{app:interaction}
The effective cavity-mediated interaction is fixed by the mode structure of the cavity and factorizes into a transversal and longitudinal part,
\begin{equation}
    D^\mathrm{3D}(\mathbf{r}, \mathbf{r}', z, z') = \sum_\mu \frac{\mathcal{U}_\mu(\mathbf{r}) \mathcal{U}_\mu(\mathbf{r}')}{\Delta_{c,\mu}/\Delta_{c,0}} = D(\mathbf{r}, \mathbf{r}')\,\cos(z)\cos(z'),
\end{equation}
where $\mathbf{r}=(x,y)$ denotes the transverse coordinate and we use the multi-index $\mu = (m, l)$. The transverse interaction is the weighted sum over the normalized and orthonormal Hermite-Gaussian cavity modes
\begin{equation}\label{eq:D-def}
    D(\mathbf{r}, \mathbf{r}') = \sum_\mu \frac{u_\mu(\mathbf{r})\, u_\mu(\mathbf{r}')}{\Delta_{c,\mu}/\Delta_{c,0}},
\end{equation}
where the transverse modes of the near planar cavity are given by
\begin{equation}
    u_{ml}(\mathbf{r}) =  \left(\frac{2}{\pi}\right)^{1/2} \frac{1}{\sqrt{2^{m+l} m!\,l!}\,w_0}\, H_m\!\left(\frac{\sqrt{2}\,x}{w_0}\right) H_l\!\left(\frac{\sqrt{2}\,y}{w_0}\right) e^{- \frac{x^2+y^2}{w_0^2}}.
\end{equation}
Here, $H_m$ denotes the Hermite polynomial of order $m$ and $w_0$ is the beam waist. We assume an exponential mode dispersion $\Delta_{c,\mu} = \Delta_c\, e^{n_\mu\epsilon}$, where $n_\mu = l + m$ denotes the order of the transverse mode and $\epsilon = \tilde \epsilon/\Delta_c \ll 1$ quantifies the departure from ideal mode degeneracy. This quantity can be related to the number of modes $M^* \approx 1/\epsilon$ participating in the interaction and $\tilde \epsilon$ is related to the cavity's length deviation from the ideal planar case \cite{vaidyaTunableRangePhotonMediatedAtomic2018b}.  For small $n_\mu$ the exponential dispersion reduces to the linear one used in \cite{guoOpticalLatticeSound2021, guoSignChangingPhotonMediatedAtom2019, vaidyaTunableRangePhotonMediatedAtomic2018b} but provides a regularization of the interaction kernel at small separations.
Since both the modes and the dispersion weight factorize over the two Cartesian directions, $u_{ml}(\mathbf{r}) = u_m(x)\,u_l(y)$ and $e^{-n_\mu\epsilon / \Delta_c} = e^{-m\epsilon / \Delta_c}e^{-l\epsilon /\Delta_c}$ with $\mu=(m,l)$, the transverse kernel splits as
\begin{equation}\label{eq:D-factorized}
    D(\mathbf{r},\mathbf{r}') = K(x,x')\,K(y,y'),
    \qquad
    K(x,x') = \sum_n e^{- n\epsilon/\Delta_c}\, u_n(x)\,u_n(x'),
\end{equation}
where the one-dimensional mode is $u_n(x) = \left(2/\pi w_0^2\right)^{1/4} \big(2^n n!\big)^{-1/2} H_n(\sqrt{2}\,x/w_0)\, e^{-x^2/w_0^2}$.

The one-dimensional sum can be evaluated using Mehler's formula \cite[10.13.22]{bateman_1953_cnd32-h9x80}\cite{mehler1866, stone2009mathematics}
\begin{equation}
    \sum_n e^{-\epsilon n}\frac{1}{2^n n!} H_n(X) H_n(Y)\, e^{-(X^2+ Y^2)/2} =\frac{1}{\sqrt{1 - e^{-2\epsilon}}} \exp\!\left[ \frac{4XY - (X^2 + Y^2)(1+e^{-2\epsilon})}{2(1-e^{-2\epsilon})} \right],
\end{equation}
which, in terms of the Hermite-Gaussian modes directly yields the one-dimensional kernel
\begin{equation}\label{eq:mehler}
    K(x,x') = \sum_n e^{-\epsilon/\Delta_c n}\, u_n(x)\, u_n(x') = \frac{1}{\sqrt{\pi}\,w_0}\, \frac{e^{\epsilon/2}}{\sqrt{\sinh\epsilon}}\, \exp\!\left[ -\frac{(x-x')^2}{2w_0^2\tanh(\epsilon/2)} - \frac{(x+x')^2}{2w_0^2\coth(\epsilon/2)} \right],
\end{equation}
where the hyperbolic form follows from $1-e^{-2\epsilon} = 2e^{-\epsilon}\sinh\epsilon$ and resolving the quadratic form along the relative and center-of-mass coordinates. This expression can be verified easily by going in the reverse direction.

The product \eqref{eq:D-factorized} over both directions then gives the exact transverse interaction kernel for arbitrary values of $\epsilon$
\begin{equation}\label{eq:D-exact}
    D(\mathbf{r},\mathbf{r}') = \frac{1}{\pi w_0^2}\, \frac{e^{\epsilon}}{\sinh\epsilon}\, \exp\!\left[ -\frac{|\mathbf{r}-\mathbf{r}'|^2}{2w_0^2\tanh(\epsilon/2)} - \frac{|\mathbf{r}+\mathbf{r}'|^2}{2w_0^2\coth(\epsilon/2)} \right].
\end{equation}
In the near-degenerate regime $\epsilon\ll1$ the hyperbolic functions simplify, using $\tanh(\epsilon/2)\to\epsilon/2$, $\coth(\epsilon/2)\to 2/\epsilon$, $\sinh\epsilon\to\epsilon$ and $\exp(\epsilon) \to 1$, which yields
\begin{equation}\label{eq:D-leading}
    D(\mathbf{r},\mathbf{r}') = \frac{1}{\pi w_0^2\epsilon}\,
    \exp\!\left[ -\frac{|\mathbf{r}-\mathbf{r}'|^2}{w_0^2\epsilon} - \frac{\epsilon\,|\mathbf{r}+\mathbf{r}'|^2}{4w_0^2} \right].
\end{equation}
The translational invariant term varies on the scale $w_0\sqrt{\epsilon/2}$, whereas the center of mass term varies on the much larger scale $w_0\sqrt{2/\epsilon}$. For small enough $\epsilon$ or for a system restricted to be near the center of the transverse plane this contribution can be neglected. Dropping it leaves a translationally invariant Gaussian
\begin{equation}
    D(\mathbf{r}, \mathbf{r}') = \frac{1}{\pi w_0^2 \epsilon}\, e^{-|\mathbf{r} - \mathbf{r}'|^2/w_0^2\epsilon} \equiv \frac{1}{2\pi \sigma^2}\, e^{-|\mathbf{r} - \mathbf{r}'|^2/2\sigma^2},
\end{equation}
with the interaction width $\sigma = \sqrt{w_0^2\epsilon/2}$.
This kernel is normalized, $\int d^2r'\, D(\mathbf{r},\mathbf{r}') = 1$, so that $D(\mathbf{r},\mathbf{r}') \to \delta^{(2)}(\mathbf{r}-\mathbf{r}')$ as $\epsilon \to 0$, in the limit of the ideally degenerate cavity.

\section{Bogoliubov Formalism}
Small fluctuations around the macroscopic mean-field solution can be described by decomposing the atomic field operator $\hat{\Psi}(\mathbf{r})$ into a condensate component and a fluctuating part:
\begin{equation}
\hat \Psi (\mathbf{r}) = \phi_0(\mathbf{r})\, \hat a_0 + \delta \hat{ \Psi} (\mathbf{r}).
\end{equation}
In the limit of a large particle number $N$, the Bogoliubov approximation replaces $\hat{a}_0$ with the c-number $\sqrt{N_0} \approx \sqrt{N}$. Since the background profile $\phi_0$ satisfies the stationary Gross-Pitaevskii equation, linear terms in $\delta \hat{\Psi}$ vanish. Assuming purely local interactions ($U_\mathrm{nl} = 0$), the atomic Hamiltonian expanded to quadratic order in the fluctuations is
\begin{equation}
\hat H_\mathrm{A} \approx E_0 + \frac{1}{2} \int d\mathbf{r} \begin{bmatrix} \delta \hat \Psi^\dagger & \delta \hat \Psi \end{bmatrix} \mathcal{L} \begin{bmatrix} \delta \hat \Psi \\ \delta \hat \Psi^\dagger \end{bmatrix},
\end{equation}
where the Bogoliubov-de Gennes (BdG) operator is
\begin{equation}
\mathcal{L} = \begin{bmatrix} \hat{H}_{GP} + U_0 |\phi_0|^2 & U_0 \phi_0^2 \\ -U_0 (\phi_0^*)^2 & -(\hat{H}_{GP}^* + U_0 |\phi_0|^2) \end{bmatrix},
\end{equation}
with the Gross-Pitaevskii operator $\hat{H}_{GP} = -\frac{\hbar^2}{2M}\nabla^2 + V_{\text{eff}}(\mathbf{r}) + U_0 |\phi_0|^2 - \mu$.

For a homogeneous system of $N$ particles where the $\mathbf{k}=0$ mode is macroscopically occupied, we expand the fluctuations in a plane-wave basis: $\delta \hat \Psi(\mathbf{r}) = \sum_{\mathbf{k}\ne0} \hat a_{\mathbf{k}} e^{i \mathbf{k} \cdot \mathbf{r}}$. In momentum space, the BdG matrix couples particle pairs $(\mathbf{k}, -\mathbf{k})$:
\begin{equation}
\mathcal{L}_{\mathbf{k}} =
\begin{bmatrix}
\epsilon_{\mathbf{k}} + n_0 U_0 & n_0 U_0 \\
-n_0 U_0 & - (\epsilon_{\mathbf{k}} + n_0 U_0)
\end{bmatrix},
\end{equation}
where $n_0 = |\phi_0|^2$ and $\epsilon_{\mathbf{k}} = \frac{\hbar^2 \mathbf{k}^2}{2M}$ is the free-particle kinetic energy. Diagonalizing $\mathcal{L}_{\mathbf{k}}$ yields the Bogoliubov dispersion relation:
\begin{equation}
E_{\mathbf{k}} = \sqrt{\epsilon_{\mathbf{k}} \left( \epsilon_{\mathbf{k}} + 2 n_0 U_0 \right)}.
\end{equation}

The Hamiltonian is diagonalized by introducing quasiparticle operators $\hat b_{\mathbf{k}}$ and $\hat b_{\mathbf{k}}^\dagger$ via the canonical Bogoliubov transformation:
\begin{equation}
\begin{bmatrix} \hat a_{\mathbf{k}} \\ \hat a_{-\mathbf{k}}^\dagger \end{bmatrix}
=
\begin{bmatrix} u_{\mathbf{k}} & v_{\mathbf{k}} \\ v_{\mathbf{k}} & u_{\mathbf{k}} \end{bmatrix}
\begin{bmatrix} \hat b_{\mathbf{k}} \\ \hat b_{-\mathbf{k}}^\dagger \end{bmatrix},
\end{equation}
with real coefficients
\begin{equation}
u_{\mathbf{k}}
=\sqrt{\frac{\epsilon_{\mathbf{k}}+U_0 n_0}{2E_{\mathbf{k}}}+\frac{1}{2}},
\qquad
v_{\mathbf{k}}
=-\sqrt{\frac{\epsilon_{\mathbf{k}}+U_0 n_0}{2E_{\mathbf{k}}}-\frac{1}{2}},
\end{equation}
which satisfy the canonical normalization condition $u_{\mathbf{k}}^2-v_{\mathbf{k}}^2=1$.
This yields the diagonalized atomic Hamiltonian (omitting constant zero-point contributions):
\begin{equation}
\hat H_A = \sum_{\mathbf{k}\neq 0} E_{\mathbf{k}} \hat b_{\mathbf{k}}^\dagger \hat b_{\mathbf{k}}.
\end{equation}

Using the polar decomposition $\hat{\Psi} = \sqrt{n}\, e^{i\theta}$, the fluctuation operator is expressed in terms of canonically conjugate density and phase fluctuations, $[\delta \hat n(\mathbf r), \delta \hat\theta(\mathbf r')] = i\delta(\mathbf r-\mathbf r')$:
\begin{equation}
\delta \hat\Psi(\mathbf r) = \frac{1}{2\sqrt{n_0}} \delta \hat n(\mathbf r) + i\sqrt{n_0}\,\delta \hat\theta(\mathbf r).
\end{equation}
In the phonon regime ($\epsilon_{\mathbf k} \ll 2U_0 n_0$), the dispersion becomes linear $E_{\mathbf k} \simeq \hbar c |\mathbf k|$ with sound velocity $c = \sqrt{U_0n_0/M}$. Defining $\Lambda_{\mathbf k}=U_0n_0/E_{\mathbf k}\gg1$, the coefficients are approximated as
\begin{equation}
u_{\mathbf k}\simeq\frac{1}{\sqrt2}\sqrt{\Lambda_{\mathbf k}+1},
\qquad
v_{\mathbf k}\simeq-\frac{1}{\sqrt2}\sqrt{\Lambda_{\mathbf k}-1}.
\end{equation}
The $\pm1$ terms are kept in this final approximation so that the Bogoliubov normalization condition $u_{\mathbf k}^2-v_{\mathbf k}^2=1$ remains exactly satisfied. Consequently,
\begin{equation}
u_{\mathbf k}-v_{\mathbf k}\simeq\sqrt{2\Lambda_{\mathbf k}},
\qquad
u_{\mathbf k}+v_{\mathbf k}\simeq\frac{1}{\sqrt{2\Lambda_{\mathbf k}}},
\end{equation}
which gives the density and phase fluctuations
\begin{align}
\delta \hat n(\mathbf r) &\approx \frac{1}{\sqrt V} \sum_{\mathbf k} \sqrt{\frac{E_{\mathbf k}}{2U_0}} \left( \hat b_{\mathbf k} e^{i\mathbf k\cdot\mathbf r} + \hat b_{\mathbf k}^\dagger e^{-i\mathbf k\cdot\mathbf r} \right), \\
\delta \hat\theta(\mathbf r) &\approx -\frac{i}{\sqrt V} \sum_{\mathbf k} \sqrt{\frac{U_0}{2E_{\mathbf k}}} \left( \hat b_{\mathbf k} e^{i\mathbf k\cdot\mathbf r} - \hat b_{\mathbf k}^\dagger e^{-i\mathbf k\cdot\mathbf r} \right).
\end{align}

\section{Detector Response Rate and the Role of the Classical Drive}
\label{app:response_rate}

The cavity-mediated interaction couples the detector to the smeared density of the BEC. In the interaction picture, parametrized by the detector's proper time $\tau$, the interaction Hamiltonian reads
\begin{equation}
    \hat H_I(\tau) = \hbar\lambda\,\chi(\tau)\,\hat n_\sigma(\tau)
    \left(\hat\sigma_+\, e^{-i\Delta_\mathrm{D}\tau}
        + \hat\sigma_-\, e^{+i\Delta_\mathrm{D}\tau}\right),
    \label{eq:HI}
\end{equation}
where $\lambda$ is the coupling strength, $\Delta_\mathrm{D}=E_e-E_g$ is the detector gap, $\hat\sigma_\pm$ are the detector transition operators, and
\begin{equation}
    \hat n_\sigma(\tau) = \int \mathrm{d}^d r\,
    D(\mathbf{r},\mathbf{r}_\mathrm{D})\,
    \hat\phi^\dagger(\mathbf{r},\tau)\hat\phi(\mathbf{r},\tau)
    \label{eq:smeared_density}
\end{equation}
is the condensate density smeared over the detector profile $D(\mathbf{r},\mathbf{r}_\mathrm{D})$. The factor $\chi(\tau)$ is an adiabatic switching function. In the following, we choose $\chi(\tau)=e^{-s|\tau|}$ and take $s\to0$ at the end of the calculation, although its exact form is not relevant as long as $s$ is small.

Expanding the field around the uniform and static condensate ground state, $ \hat\phi=\sqrt{n_0}+(\delta\hat\varphi_\mathrm{re}+i\delta\hat\varphi_\mathrm{im})/\sqrt{2}$, and keeping terms up to first order in the fluctuations, the smeared density \eqref{eq:smeared_density} splits into a background and a fluctuation part,
\begin{equation}
    \hat n_\sigma(\tau) = n_0 + \delta\hat n_\sigma(\tau),
    \qquad
    \delta\hat n_\sigma(\tau) = \sqrt{2n_0}\int \mathrm{d}^d r\,
    D(\mathbf{r},\mathbf{r}_\mathrm{D})\,
    \delta\hat\varphi_\mathrm{re}(\mathbf{r},\tau).
    \label{eq:density_split}
\end{equation}
The interaction Hamiltonian therefore separates into a classical drive, governed by the coupling to the background density $n_0$, and a coupling to the smeared density fluctuation $\delta\hat n_\sigma$, which realizes the effective Unruh--DeWitt detector model of the main text. In the regime of interest the relevant density excitations are long-wavelength phonons, whose Bogoliubov dispersion linearizes. The density fluctuations $\delta\hat n_\sigma\propto\hat\varphi_\mathrm{re}$ then behave as a massless relativistic scalar field on an emergent acoustic spacetime, with the speed of sound $c_s$ replacing the light speed. Its vacuum is stationary, so one-point functions vanish and correlators depend only on the proper-time difference. 

We now show that the classical drive does not contribute to the detector response rate. The response rate, the transition probability per unit of proper time, is given by Fermi's golden rule,
\begin{equation}
    F = \lim_{s\to0}\lim_{\tau_0\to\infty}\frac{1}{2\tau_0}
    \sum_n\left|\frac{i}{\hbar}\int_{-\tau_0}^{\tau_0}\mathrm{d}\tau\,
    \langle\psi_n,e|\hat H_I(\tau)|0,g\rangle\right|^2 ,
    \label{eq:FGR}
\end{equation}
where the initial state $|0,g\rangle=|0\rangle\otimes|g\rangle$ is a product state between the field
vacuum times the detector ground state, and the final state is given by the detector being in its excited state $|\psi_n,e\rangle=|\psi_n\rangle\otimes|e\rangle$ with $|\psi_n\rangle$ an arbitrary final field configuration. Using \eqref{eq:HI} and \eqref{eq:density_split}, the transition amplitude becomes
\begin{equation}
    c_n = -i\lambda\int_{-\tau_0}^{\tau_0}\mathrm{d}\tau\,\chi(\tau)\,
    e^{-i\Delta_\mathrm{D}\tau}
    \Big(n_0\,\langle\psi_n|0\rangle
        + \langle\psi_n|\delta\hat n_\sigma(\tau)|0\rangle\Big).
\end{equation}
Its modulus squared contains three contributions,
\begin{equation}
\begin{split}
    |c_n|^2 = \lambda^2\!\int_{-\tau_0}^{\tau_0}\!\!\mathrm{d}\tau
    \int_{-\tau_0}^{\tau_0}\!\!\mathrm{d}\tau'\,
    \chi(\tau)\chi(\tau')\,e^{-i\Delta_\mathrm{D}(\tau-\tau')}
    \Big[
        &\,n_0^2\,\langle0|\psi_n\rangle\langle\psi_n|0\rangle \\
        &+ \langle0|\delta\hat n_\sigma(\tau')|\psi_n\rangle
           \langle\psi_n|\delta\hat n_\sigma(\tau)|0\rangle \\
        &+ n_0\,\langle0|\psi_n\rangle
           \langle\psi_n|\delta\hat n_\sigma(\tau)|0\rangle
        + \mathrm{H.c.}
    \Big],
\end{split}
\end{equation}
which correspond, respectively, to transitions driven by the classical
background, to excitations caused by the coupling to the condensate
fluctuations, and to the cross terms between the two. Summing over all final
field states and inserting the completeness relation
$\sum_n|\psi_n\rangle\langle\psi_n|=\mathds{1}$ gives
\begin{equation}
\begin{split}
    \sum_n|c_n|^2 = \lambda^2\!\int_{-\tau_0}^{\tau_0}\!\!\mathrm{d}\tau
    \int_{-\tau_0}^{\tau_0}\!\!\mathrm{d}\tau'\,
    \chi(\tau)\chi(\tau')\,e^{-i\Delta_\mathrm{D}(\tau-\tau')}
    \Big[
        &\,n_0^2
        + \langle0|\delta\hat n_\sigma(\tau')\,\delta\hat n_\sigma(\tau)|0\rangle \\
        &+ n_0\,\langle0|\delta\hat n_\sigma(\tau)|0\rangle + \mathrm{H.c.}
    \Big].
\end{split}
\end{equation}
Since the one-point function of the fluctuation field vanishes,
$\langle0|\delta\hat n_\sigma(\tau)|0\rangle=0$, the cross terms drop out, and
the response separates into the sum of a classical-drive and a fluctuation
contribution,
\begin{equation}
    \sum_n|c_n|^2 = \lambda^2\!\int_{-\tau_0}^{\tau_0}\!\!\mathrm{d}\tau
    \int_{-\tau_0}^{\tau_0}\!\!\mathrm{d}\tau'\,
    \chi(\tau)\chi(\tau')\,e^{-i\Delta_\mathrm{D}(\tau-\tau')}
    \Big[n_0^2
        + \langle0|\delta\hat n_\sigma(\tau')\,\delta\hat n_\sigma(\tau)|0\rangle
    \Big].
    \label{eq:two_contributions}
\end{equation}
For the classical contribution, the double integral factorizes and requires no
adiabatic regulator. Taking the coupling to be constant over the interaction
interval, $\chi=1$, gives
\begin{equation}
    \sum_n|c_n|^2_\mathrm{cl}
    = \lambda^2 n_0^2\left|\int_{-\tau_0}^{\tau_0}\mathrm{d}\tau\,
        e^{-i\Delta_\mathrm{D}\tau}\right|^2
    = \lambda^2 n_0^2\,\frac{4\sin^2(\Delta_\mathrm{D}\tau_0)}{\Delta_\mathrm{D}^2}
    \le \frac{4\lambda^2 n_0^2}{\Delta_\mathrm{D}^2} ,
\end{equation}
which oscillates with $\tau_0$ and remains bounded rather than growing with
time. The classical drive therefore produces only Rabi oscillations---a
bounded, reversible admixture of the excited state of order
$P_\mathrm{cl}\sim(\lambda n_0/\Delta_\mathrm{D})^2$---rather than an
accumulating transition probability. Consequently, its long-time rate
vanishes,
\begin{equation}
    F_\mathrm{cl}
    = \lim_{\tau_0\to\infty}\frac{1}{2\tau_0}\,
      \lambda^2 n_0^2\,\frac{4\sin^2(\Delta_\mathrm{D}\tau_0)}{\Delta_\mathrm{D}^2}
    = 0 .
\end{equation}
The same argument justifies neglecting the direct coherent drive of the
detector by the pump.
The remaining fluctuation-induced contribution in
Eq.~\eqref{eq:two_contributions} depends only on the proper-time difference.
Introducing the two-point function 
\begin{equation}
    g(\tau) = \langle0|\delta\hat n_\sigma(\tau)\,\delta\hat n_\sigma(0)|0\rangle
    = 2n_0\!\int\!\mathrm{d}^d r\!\int\!\mathrm{d}^d r'\,
        D(\mathbf{r},\mathbf{r}_\mathrm{D})
        D(\mathbf{r}',\mathbf{r}_\mathrm{D})\,
        \langle0|\hat\varphi_\mathrm{re}(\mathbf{r},\tau)
            \hat\varphi_\mathrm{re}(\mathbf{r}',0)|0\rangle ,
\end{equation}
we change variables to $u=\tau' - \tau$ and
$v=(\tau+\tau')/2$, for which $\mathrm{d}\tau\,\mathrm{d}\tau'
=\mathrm{d}u\,\mathrm{d}v$. In the long-interaction-time limit, the integral
over $v$ yields the factor $2\tau_0$ that cancels the prefactor in
Eq.~\eqref{eq:FGR}. Introducing $\omega=-\Delta_D$ the response rate reduces to the standard UdW form
\begin{equation}
    F(\omega=-\Delta_\mathrm{D})
    = \lambda^2\lim_{s\to0^+}\int_{-\infty}^{\infty}\mathrm{d}\tau\,
      e^{-i\omega\tau-s|\tau|}\,g(\tau).
    \label{eq:UdW_rate}
\end{equation}
The factor $e^{-s|\tau|}$ is the adiabatic regulator conventionally used in
the literature. Its precise form is immaterial here because it only regularizes
the integral as $|\tau|\to\infty$. When $g(\tau)$ is regular, this regulator is
redundant and is therefore omitted from our subsequent calculations.
Alternatively, one may integrate along the contour
$\tau\to\tau-i\epsilon$ and take $\epsilon\to0^+$.

\subsection{Derivation of the Response Rate for Circular Unruh Effect}
Consider a detector moving in uniform circular motion in the $x-y$ plane. Its wordline  parameterized by proper time $\tau$ is given by:
\begin{equation}
(t(\tau), \mathbf{r}(\tau)) =
(\gamma \tau, R \cos (\gamma \Omega \tau), R \sin (\gamma \Omega \tau)). 
\end{equation}
where $R$ is the orbital radius, $\Omega$ is the angular velocity, and $\gamma = 1/\sqrt{1 - (R\Omega/c)^2}$ is the Lorentz factor connecting coordinate time $t$ to proper time $\tau$.

The transition probability of the detector interacting with a scalar field vacuum involves evaluating an integral over the wave vector $\mathbf{k}$. Working in polar coordinates $\mathbf{k} = (k\cos\phi, k\sin\phi)$, we introduce a Gaussian spatial smearing profile $e^{-\sigma^2 k^2}$ of width $\sigma$ to regulate ultraviolet divergences. The response integral in the time domain is:
\begin{equation}
I(\tau) = \int_0^\infty \int_0^{2\pi} e^{-\sigma^2 k^2} k^2 e^{i (\mathbf{k} \cdot {\Delta \mathbf{r}} - c k \gamma \tau)} \, d\phi \, dk
\end{equation}

To calculate the integral, we first compute the dot product between the wave vector and the trajectory:
\begin{align}
\mathbf{k} \cdot {\Delta \mathbf{r}}= kR \left[ \cos \phi (\cos(\gamma \Omega \tau) - 1) + \sin \phi \sin(\gamma \Omega \tau) \right]  = kR \left[ \cos(\phi - \gamma \Omega \tau) - \cos \phi \right] \\ = 2kR \sin\left(\frac{\gamma \Omega \tau}{2}\right) \sin\left(\phi - \frac{\gamma \Omega \tau}{2}\right)
\end{align}
Using the identity
\begin{equation}
\int_0^{2\pi} e^{i x \cos(\phi - \alpha)} d\phi = 2\pi J_0(x),
\end{equation}
where $J_0(x)$ is the zeroth-order Bessel function of the first kind, 
the inner angular integral yields:
\begin{equation}
I(\tau) = 2\pi \int_0^\infty k^2 e^{-\sigma^2 k^2} e^{-i c \gamma \tau k} J_0\left(d(\tau) k\right) dk,
\quad \mbox{ with }
\quad d(\tau) = 2R \sin\left(\frac{\gamma \Omega \tau}{2}\right).
\end{equation}

Using the Bessel-function identity employed in analyses of circular detector response \cite{Louko2018}
\begin{equation}
    J_0\left( 2 a \sin x \right) = \sum_{n=-\infty}^{\infty} J_n^2(a) e^{2 i n x},
\end{equation}
 we find
 \begin{align}
\mathcal{F}(\omega) \propto \int_{-\infty}^{\infty} I(\tau) e^{-i \omega \tau} d\tau  =  2\pi \sum_n \int_{-\infty}^{\infty} d\tau\int_0^\infty  dk\,k^2 e^{-\sigma^2 k^2} e^{-i c \gamma \tau k} e^{-i \omega \tau} 
J_n^2\left(R k\right) e^{i n \gamma \Omega \tau}
 \\
=2\pi \sum_n \int_0^\infty  dk\,k^2 e^{-\sigma^2 k^2} 
J_n^2\left(R k\right) \underbrace{\left[\int_{-\infty}^{\infty} d\tau e^{i n \gamma \Omega \tau} e^{-i c \gamma \tau k} e^{-i \omega \tau} 
\right]}_{2\pi \delta(\omega - n\gamma\Omega + ck\gamma) = \frac{2\pi}{c \gamma} \delta  \left( k -\frac{(n\gamma\Omega-\omega )}{c \gamma}\right)}   = \frac{4\pi^2}{c \gamma} \sum_n k_n^2 e^{-\sigma^2 k_n^2} 
J_n^2\left(R k_n\right),
 \end{align} 
 with $k_n = \frac{(n\gamma\Omega-\omega )}{c \gamma}$.

\section{Alternative Pumping Setup}

\begin{figure}[ht]
    \centering
    \includegraphics[trim=30 30 30 35, width=0.4\linewidth]{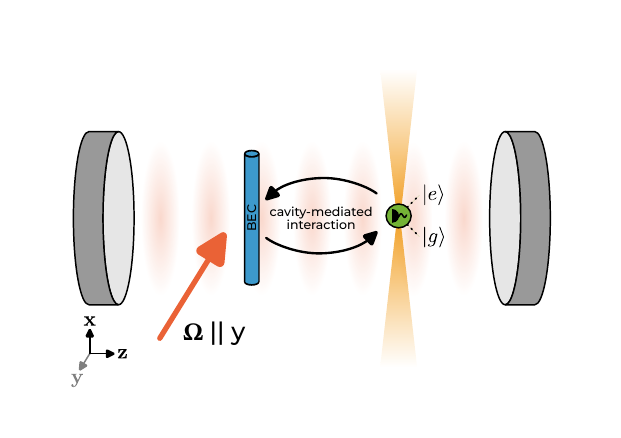}
    \caption{Alternative setup with a transversally pumped quasi-1D BEC.} 
    \label{fig:alt_setup}
\end{figure}

An alternative setup is transverse pumping, sketched in Fig.~\ref{fig:alt_setup}. Here, a one-dimensional Bose--Einstein condensate is confined in the transverse plane of a near-planar cavity and driven by a running-wave laser propagating perpendicular to the BEC and the cavity axis. Photons are then scattered from the pump into the (nearly) degenerate cavity modes by the BEC itself, so that the intracavity field is generated by the condensate rather than driven externally. This avoids the direct detector drive of the main setup. The resulting mode structure again mediates an effective finite-size interaction between the condensate and the spatially separated two-level system acts as an effective Unruh--DeWitt detector. The analysis of the preceding sections carries over with only minor modifications.

\end{document}